# Collisionless Encounters and the Origin of the Lunar Inclination

Kaveh Pahlevan* and Alessandro Morbidelli

Laboratoire Lagrange, Université Côte d'Azur, Observatoire de la Côte d'Azur, CNRS, Boulevard de l'Observatoire C.S. 34229 06304 Nice, Cedex 4, France *Corresponding author: pahlevan@oca.eu

**The Moon is generally thought to have formed from the debris ejected by the impact of a planet-sized object with the proto-Earth towards the end of planetary accretion[1,2]. Modeling of the impact process predicts that the lunar material was disaggregated into a circumplanetary disk and that lunar accretion subsequently placed the Moon in a near equatorial orbit[3-6]. Forward integration of the lunar orbit from this initial state predicts a modern inclination at least an order of magnitude smaller than the lunar value, a long-standing discrepancy known as the lunar inclination problem[7-9]. Here we show that the modern lunar orbit provides a sensitive record of gravitational interactions with Earth-crossing planetesimals not yet accreted at the time of the Moon-forming event. The excited lunar orbit can naturally be reproduced via interaction with a small quantity of mass (corresponding to 0.0075-0.015 $M_E$ eventually accreted to the Earth) carried by a few bodies, consistent with constraints and models of late accretion[10,11]. While the process has a stochastic element, the observed value of the lunar inclination is among the most likely outcomes for a wide range of parameters. The excitation of the lunar orbit is most readily reproduced via collisionless encounters with strong tidal dissipation on the early Earth. This mechanism obviates the necessity of previously proposed (but idealized) excitation mechanisms[12,13] and can constrain the planet formation context of the Moon-forming event and the pristineness of the dynamical state of the Earth-Moon system.**

The Moon-forming impact is thought to have generated a compact circumplanetary disk (within < 10 Earth radii) from which the Moon rapidly accreted. Like Saturn's rings, the proto-lunar disk is expected to become equatorial on a timescale rapid relative to its evolutionary timescale. Hence, so long as the proto-lunar material disaggregated into a disk following the giant impact, the Moon is expected to have accreted within ~1° of the Earth's equator plane[6]. Tidal evolution calculations suggest that for every degree of inclination of the lunar orbit plane relative to the Earth's equator plane at an Earth-Moon (EM) separation of 10 Earth radii ($R_E$), the current lunar orbit would exhibit ~1/2° of inclination relative to Earth's orbital plane[7-9]. The modern ~5° lunar inclination would – without external influences – translate to a ~10° inclination to Earth's equator plane at 10 $R_E$ shortly after lunar accretion. This ~10x difference between theoretical expectations of lunar accretion and the EM system has become known as the lunar inclination problem.

Previous work on this problem has sought to identify mechanisms such as a gravitational resonance between the newly formed Moon and the Sun[12] or the remnant proto-lunar disk[13] that can excite the lunar inclination to a level consistent with its current value. Neither of these scenarios is satisfactory, however, as the former requires particular values of the tidal dissipation parameters while the latter has only been shown to be viable in an idealized system where a single, fully-formed Moon interacts with a single pair of resonances in the proto-lunar disk. Moreover, prior works have all assumed that the excitation of the lunar orbit was determined during interactions essentially coinciding with lunar origin. Here, we propose that the lunar inclination arose much later as a consequence of the sweep-up of remnant planetesimals in the inner Solar System.

After the giant impact and at most ~$10^3$ years[14,15], the Moon has accreted, interacted with[13] and caused the collapse of the remnant proto-lunar disk onto the Earth[6], passed the evection resonance with the Sun[3,12,16], and begun a steady outward tidal evolution. On a timescale (~$10^6$-$10^7$ years) rapid relative to that characterizing depletion of planetesimals in the final post-Moon formation stage of planetary accretion[17] (called "late accretion"), the lunar orbit expands through the action of tides to an EM separation of ~20-40 $R_E$. In this time, the lunar orbit transitions from precession around the spin-axis of Earth to precession around the heliocentric orbit normal vector[8], and its inclination becomes insensitive to the shifting of the Earth's equatorial plane via subsequent accretion[18]. However, as we show below, lunar inclination becomes more sensitive to gravitational interactions with passing planetesimals as the tidal evolution of the system proceeds. The sensitivity is such as to render the lunar orbital excitation a natural outcome of the sweep-up of the leftovers of accretion and to yield a new constraint on the dynamical and tidal environment of the EM system in the $\sim 10^8$ years immediately following its origin.

While subsequent collisions with the EM system have been previously considered as a mechanism for dynamical excitation[18], the collision of inner Solar System bodies with the Earth tends to be preceded by a large number ($10^3$-$10^4$) of collisionless encounters. Excitation via this process is governed by two relevant timescales: the timescale over which remnant populations in the inner Solar System are lost via accretion onto the planets and/or the Sun (several tens of Myr[17]) and the timescale for the lunar tidal orbital expansion, which is a rapidly varying function of the EM distance. The EM distance is important because it determines the system cross-section for collisionless encounters with remnant bodies. The rate of tidal expansion of the lunar orbit during the first $\sim 10^8$ years after the giant impact accordingly plays an important role. As tidal evolution proceeds and the EM separation increases, the system becomes increasingly susceptible to collisionless excitation, while populations capable of exciting the system are progressively depleted. A few tens of millions of years after the Moon-forming event, the EM system reaches an optimal capacity for excitation via gravitational encounters: a dynamically excitable system co-existing with a significant remnant body population.

Here, we run a series of Monte Carlo simulations to set constraints on the outcome of repeated encounters of massive bodies with the evolving EM system. The simulations are carried out until the populations are exhausted either through collision with the Earth or through non-terrestrial loss channels (for details, see Methods). A sample run of dynamical excitation during the first ~$10^8$ years of EM history is shown in Figure 1. No single event dominates: several strong encounters contribute significantly to the final excitation. The size-distribution of the late-accreting population is taken to be top-heavy with most of the mass contained in a few massive bodies, as previously proposed to explain terrestrial late accretion[10,11]. This particular simulation ultimately results in two 0.00375 $M_E$ planetesimals left over from the formation process colliding with the Earth. Note that tidal damping of the lunar inclination is applied along with the lunar orbital expansion, following equation S1 (see Methods). We do not consider the possibility that the lunar inclination might have been more strongly damped via dissipation in the lunar magma ocean, as recently proposed[19]. In the supplementary methods section, we show that this effect is not important as long as the lunar magma ocean crystallized within a few $10^7$ years.

Lunar orbital excitation in this epoch depends on the total mass of leftover planetesimals, the number of bodies carrying the mass and their orbital distribution, the rate of terrestrial tidal dissipation, and a stochastic element. In Figure 2, we show results of simulations of the excitation of the lunar inclination due to interaction of the system with a small amount of mass (equivalent to 0.0075-0.015 $M_E$ eventually accreted to the Earth), for different values of the strength of tidal dissipation and the number of bodies delivering the mass (constrained to be <20 colliding with Earth via models of late accretion[10,11]). Several features are apparent. First, there is a quasi-linear dependence of the excitation on the total mass of late accretion: other variables being equal, excitation corresponding to 0.015 $M_E$ of late accretion is approximately twice as great as that with 0.0075 $M_E$. The mass accreted onto the Earth thereby provides a proxy for collisionless excitation. Second, the lunar orbital excitation exhibits some dependence on the strength of tidal evolution: stronger dissipation within the Earth drives the lunar orbit outwards faster and exposes the system to more collisionless events. Simulations with the weakest tides that we considered – characterized by $k_2/Q$ (=0.01) – typically excite the lunar inclination with a planetesimal population consistent with 0.015 $M_E$ of late accretion, while with stronger tidal dissipation (=0.1), lunar inclination is routinely excited by a planetesimal population carrying 0.0075 $M_E$ of late accretion. Third, there exists a negative dependence of the excitation on the number of bodies involved in late accretion, such that the mechanism requires a population that is top-heavy (with most of the mass delivered via the most massive bodies). For a given mass of late accretion, a greater number of bodies also renders the distribution of lunar inclinations more strongly peaked and predictions of the expected excitation more precise. Despite an order of magnitude of uncertainty in the strength of early terrestrial tides ($k_2/Q$) and in the number of bodies involved in the leftover population, and an intrinsic stochasticity inherent to collisionless encounters, what is striking is the robustness with which close encounters with a population of planetesimals delivering 0.0075-0.015 $M_E$ to the Earth after the Moon-forming event can reproduce the excitation that characterizes the lunar orbit.

The angular momentum of the EM system at the time of its origin is a central feature diagnostic of various giant impact scenarios proposed to date[1,3,4,18]. Given that the lunar orbit provides a sensitive dynamical measure of encounters with the EM system following its origin, we can ask whether such gravitational interactions were effective at injecting or extracting angular momentum. Figure 3 summarizes the angular momentum change versus the final excitation. The change in angular momentum corresponding to modern inclination excitation of ~5 degrees is likely a few tens of percent or less. Hence, the standard giant impact scenario[4] followed by little subsequent dynamical modification is compatible with the dynamical state of the modern system, while a high angular momentum impact scenario[3,5] would require another dynamical mechanism such as the evection resonance[3,12,16] to be reconciled with the modern EM system.

The exquisite sensitivity of the orbits of impact-generated satellites to ongoing accretion onto the host planet has several consequences. The degree of orbital excitation resulting from interaction with and accretion of 0.0075-0.015 $M_E$ onto the early Earth suggests that collisionless encounters with massive bodies – such as the lunar-to-Mars mass embryos thought to have played a key role in the accretion of the Earth – would have excited satellites to very eccentric orbits ultimately leading to their dynamical loss, either via collision with the host planet or liberation into heliocentric orbit. Such excitability of impact-generated satellite orbits may explain several outstanding features of the inner Solar System. For example, despite impact-generated satellites being a quasi-generic feature of terrestrial planet formation via giant impact, the absence of an impact-generated satellite around Venus[20] and the apparent absence of a pre-Moon terrestrial satellite[21] can be understood: any such early formed satellites would have been lost via encounters with extant planetary embryos, including perhaps the Moon-forming impactor itself. Moreover, the occurrence of the Moon-forming giant impact late in the history of Earth accretion can be understood as a necessity for its survival: even moderately earlier-generated satellites would have been readily dynamically destabilized. Just as the survival of planets depends on the surrounding stellar environment[22], the survival of an impact-generated satellite is contingent on its planet formation context at the time of origin.

**References**

1   Cameron, A. G. W. & Ward, W. R. The origin of the Moon. *Lunar Sci.* **7**, 120-122, (1976).
2   Hartmann, W. K. & Davis, D. R. Satellite-sized planetesimals and lunar origin. *Icarus* **24**, 504-515, (1975).
3   Cuk, M. & Stewart, S. T. Making the Moon from a Fast-Spinning Earth: A Giant Impact Followed by Resonant Despinning. *Science* **338**, 1047-1052, (2012).
4   Canup, R. M. & Asphaug, E. Origin of the Moon in a giant impact near the end of the Earth's formation. *Nature* **412**, 708-712, (2001).
5   Canup, R. M. Forming a Moon with an Earth-like Composition via a Giant Impact. *Science* **338**, 1052-1055, (2012).
6   Ida, S., Canup, R. M. & Stewart, G. R. Lunar accretion from an impact-generated disk. *Nature* **389**, 353-357, (1997).
7   Mignard, F. The Lunar Orbit Revisited, III. *Moon Planets* **24**, 189-207, (1981).
8   Goldreich, P. History of Lunar Orbit. *Reviews of Geophysics* **4**, 411-&, (1966).
9   Touma, J. & Wisdom, J. Evolution of the Earth-Moon System. *Astronomical Journal* **108**, 1943-1961, (1994).
10  Bottke, W. F., Walker, R. J., Day, J. M. D., Nesvorny, D. & Elkins-Tanton, L. Stochastic Late Accretion to Earth, the Moon, and Mars. *Science* **330**, 1527-1530, (2010).
11  Raymond, S. N., Schlichting, H. E., Hersant, F. & Selsis, F. Dynamical and collisional constraints on a stochastic late veneer on the terrestrial planets. *Icarus* **226**, 671-681, (2013).
12  Touma, J. & Wisdom, J. Resonances in the early evolution of the earth-moon system. *Astronomical Journal* **115**, 1653-1663, (1998).
13  Ward, W. R. & Canup, R. M. Origin of the Moon's orbital inclination from resonant disk interactions. *Nature* **403**, 741-743, (2000).
14  Thompson, C. & Stevenson, D. J. Gravitational-Instability in 2-Phase Disks and the Origin of the Moon. *Astrophysical Journal* **333**, 452-481, (1988).
15  Salmon, J. & Canup, R. M. Lunar Accretion from a Roche-Interior Fluid Disk. *Astrophysical Journal* **760**, (2012).
16  Wisdom, J. & Tian, Z. L. Early evolution of the Earth-Moon system with a fast-spinning Earth. *Icarus* **256**, 138-146, (2015).
17  Morbidelli, A., Marchi, S., Bottke, W. F. & Kring, D. A. A sawtooth-like timeline for the first billion years of lunar bombardment. *Earth and Planetary Science Letters* **355**, 144-151, (2012).
18  Canup, R. M. Dynamics of lunar formation. *Annual Review of Astronomy and Astrophysics* **42**, 441-475, (2004).
19  Nimmo, F. & Chen, E. in *Lunar and Planetary Science Conference XLV*   (The Woodlands, Texas, 2014).
20  Alemi, A. & Stevenson, D. J. Why Venus has No Moon. *Bulletin of the American Astronomical Society* **38**, 491, (2006).

21  Canup, R. M. Lunar-forming impacts: processes and alternatives. *Philosophical Transactions of the Royal Society a-Mathematical Physical and Engineering Sciences* **372**, (2014).
22  Spurzem, R., Giersz, M., Heggie, D. C. & Lin, D. N. C. Dynamics of Planetary Systems in Star Clusters. *Astrophysical Journal* **697**, 458-482, (2009).

**Acknowledgements**

This research was carried out as part of a Henri Poincaré Fellowship at the Observatoire de la Côte d'Azur (OCA) to K.P. The Henri Poincaré Fellowship is funded by the OCA and the City of Nice, France. A.M. would like to thank the European Research Council Advanced Grant ACCRETE (#290568).

**Author Contributions**

K.P. and A.M. discussed every step of the project, designed the simulation set-up and co-wrote the numerical code. K.P. performed the simulations and the statistical analysis.

**Author Information**

The authors declare no competing financial interests. Correspondence should be addressed to pahlevan@oca.eu.

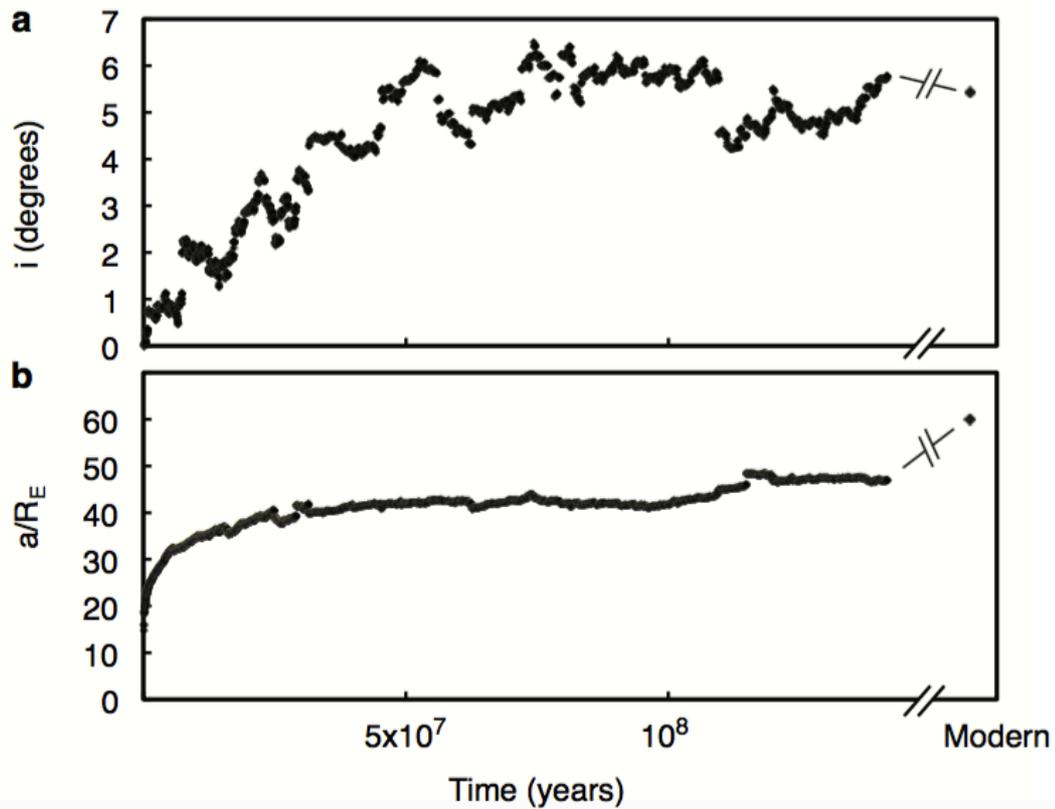

**Figure 1** – Sample realization – A model of the early lunar orbit subject to tidal evolution ($k_2/Q=0.1$) and encounters leading to collision of two 0.00375 $M_E$ bodies with the Earth. The semi-major axis of the evolving lunar orbit is given in Earth radii. While not every encounter increases the lunar inclination, the cumulative effect is a tendency toward excitation. Notable interactions include merging collisions with the Earth kicking the lunar orbit via recoil (at 29.1 and 31.5 Myrs), several exceptionally close encounters with the Moon (at 7.3 and 109.6 Myrs) and the exhaustion of the population (at 141.7 Myrs) ultimately marking the end of the simulation. Note that subsequent inclination damping via planetary tides is modest (from 5.8° at 47 $R_E$ at the end of the simulation to 5.4° at 60 $R_E$), a feature that is typical of this "late" excitation mechanism.

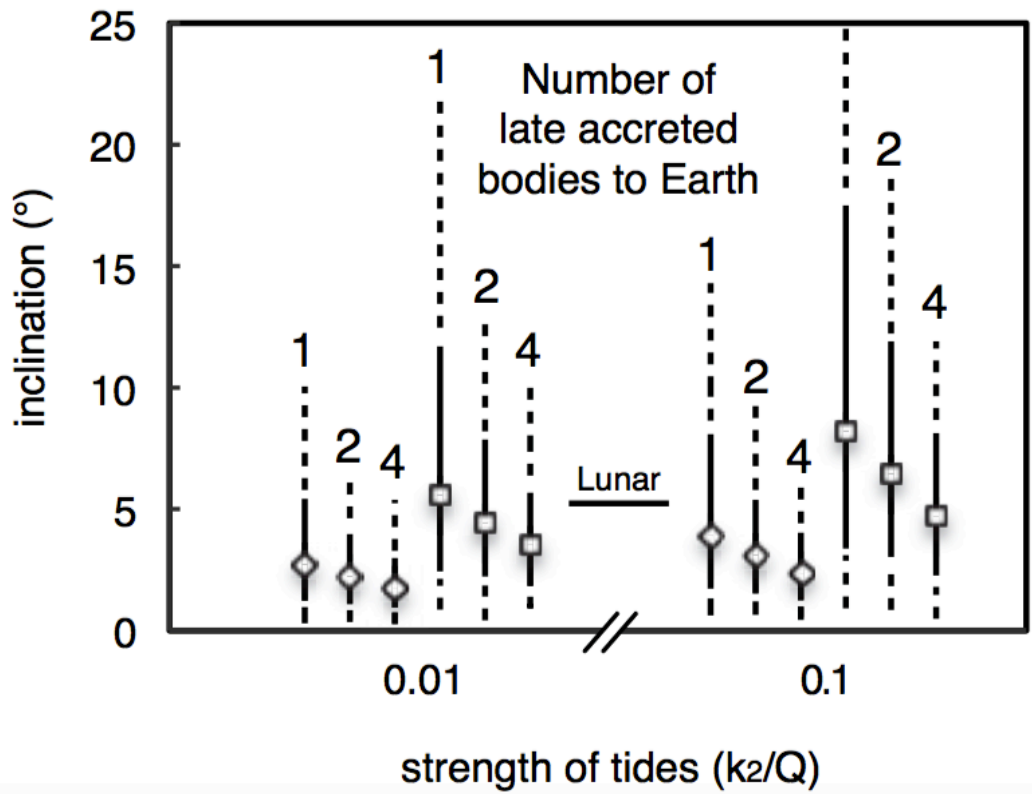

**Figure 2** – Summary of simulations – Median values, 1σ (solid) and 2σ (dashed) intervals for the lunar inclination at the end of the simulations after damping via planetary tides to the modern Earth-Moon separation. The excitation in the modern lunar orbit is plotted for comparison. Diamonds correspond to simulations with 0.0075 $M_E$ accreted to Earth, squares to 0.015 $M_E$. The "strong" tidal dissipation case ($k_2/Q=0.1$) corresponds to a hot dissipative silicate Earth while the "weak" dissipation ($k_2/Q=0.01$) represents the geologic average value dominated by dissipation in shallow oceans. In these simulations, the accretion of 0.0075 $M_E$ (with "strong" tides) to 0.015 $M_E$ (with "weak" tides) frequently reproduces the excitation in the lunar orbit. The number of bodies delivering the late accreted mass in each set of simulations is reported over each symbol.

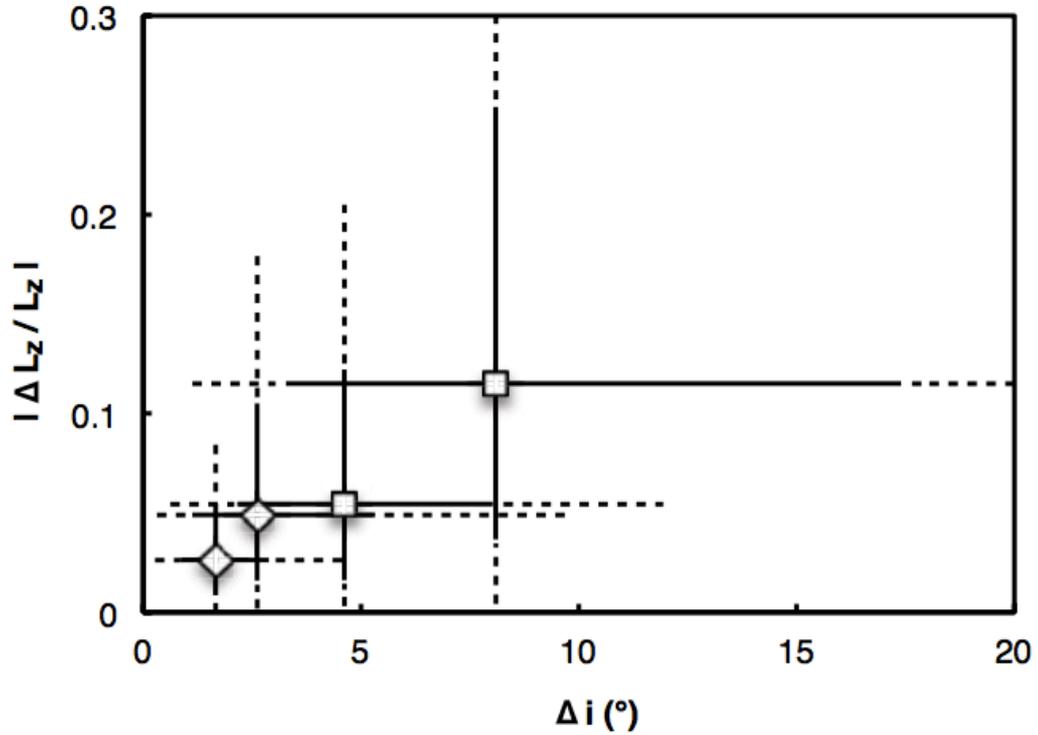

**Figure 3** – Angular momentum change of the EM system – Median values, 1σ (solid) and 2σ (dashed) intervals are shown for inclination and angular momentum change via post-lunar collisionless encounters. Diamonds represent realizations with weak tides ($k_2/Q=0.01$) and 0.0075 $M_E$ accretion while squares correspond to cases with stronger dissipation ($k_2/Q=0.1$) and 0.015 $M_E$ accretion, bracketing the range in our simulations. Each suite of simulations is composed of two subsets: one with late accretion delivered via 1 body (greater excitation), the other 4 (lesser excitation). Intermediate outcomes with 2 accreted bodies are omitted for clarity. For a level of excitation consistent with the modern lunar orbit (5.15°), the amount of system angular momentum change is likely < 20%. However, the 2σ intervals for the strongest excitation case plotted extends to $\Delta i$ = 42° and $|\Delta L_z/L_z|$ = 0.48 implying a small probability (<5%) for angular momentum change >50%.

**Methods**

A large number of simulations (~$10^3$) are required to characterize the distribution of outcomes for repeated encounters of a given planetesimal population with a given early EM system. Accordingly, we design a numerical experiment that captures the physics of the problem statistically and that can be computed efficiently. Heliocentric orbits for late accreting planetesimal populations were generated according to a Rayleigh distribution with a Rayleigh eccentricity ($e_R$=0.3) and inclination ($i_R$=$e_R$/2); these values are consistent with simulations of terrestrial planet formation[23]. To test for sensitivity to population orbits, we varied $e_R$ between 0.3-0.4 with the resulting median inclination excitation changing by less than 10%. With given orbital distributions, the subset of the population that is Earth-crossing was selected and encounter probabilities with the Hill radius ($R_H$) of the Earth were calculated according to the expressions given in the literature[24]. The masses of the planetesimals are assumed to be in the range 0.15-1.2 lunar masses, consistent with those expected for the projectiles carrying the Earth's late accretion[10]. At the beginning of the simulations, the Earth and Moon were placed on circular uninclined orbits at 1 AU and 5 $R_E$, respectively, near their orbits at the end of accretion and the beginning of tidal history. An encounter time and encounter orbit were chosen randomly according to the distribution of Earth-crossing planetesimal encounter probabilities. At the time of each encounter, phases for the lunar orbit, characterized by the argument of perigee ($\omega$), the longitude of the ascending node ($\Omega$), and the mean anomaly (M) were selected randomly, as was the orientation of the planetesimal orbit within those admitted by the selected orbital parameters. The impact parameter (b) was selected in the interval [0, $R_H$] according to a uniform encounter probability per unit area ($dP \propto b\,db$). Gravitational three-body (Earth-Moon-planetesimal) encounters were integrated with a Bulirsch-Stoer integrator included in the SWIFT package in a geocentric reference frame, tracking changes to the lunar orbit. In between 3-body encounters, the eccentricity and semi-major axis of the lunar orbit were evolved with a constant Q tidal model[25] while the lunar inclination was evolved with a model[26] for planetary tides:

$$\frac{di}{i} = -\frac{1}{4}\frac{da}{a} \qquad (M1)$$

Impacts with the Earth were assumed to be inelastic merging events with the final body carrying the total mass and momentum. Impacts onto the Moon would have been in the erosive and/or catastrophic disruption regime and realizations with such events were removed from the subsequent analysis (discussed below).

Remnant planetesimals can, in general, be eliminated via collision with the terrestrial planets and the Sun or dynamical ejection from the inner Solar System[11]. We characterize such losses using the outcome of direct N-body simulations that trace the evolution of such early planetesimals, yielding a 10-fold depletion of the Earth-crosser population in the first 100 Myrs corresponding to a population decay law of ~exp(-t/$\tau_{ss}$) with $\tau_{ss} \approx$ 44 Myrs[17]. While the modern near Earth object (NEO) population is resupplied by the asteroid belt in quasi-steady state fashion and effectively does not decay, the leftover planetesimal population is not resupplied by a larger reservoir and therefore does decay. Note that, due to partial resupply of Earth-crossing bodies, the timescale for the decay of the Earth-crossing population can nevertheless be different than the lifetime of individual particles. In order to integrate the decay rates of Earth-crossing planetesimals with our simulations, we use the following procedure: after generating orbital populations but before running three-body integrations, we allow the Earth-crossing populations to encounter the Earth alone, permitting derivation of a time constant for decay of this population solely via collision with the Earth ($\tau_E$=79 Myrs). Next, we require that the Earth-crossing planetesimal population in our three-body simulations decay at the same average rate as that observed in the N-body heliocentric simulations. We therefore decompose average loss rates of the Earth-crossing population into terrestrial and non-terrestrial loss modes ($1/\tau_{SS} = 1/\tau_E + 1/\tau_{NE}$) and thereby derive a time constant ($\tau_{NE}$=99 Myrs) for removal via non-terrestrial loss channels. Accordingly, we stochastically remove bodies from the population in our three-body simulations such that the average loss rate of Earth-crossing planetesimals – either through collision with Earth (explicit) as well as through other modes of loss (implicit) – is consistent with the average loss rates observed in N-body simulations of late accretion[17] (see Extended Data Figure 1a).

Each data point in Figure 2 results from 4000 realizations. To analyze the results, certain realizations were eliminated. These fall into three categories: (1) Collision of a planetesimal with the Moon. Occasionally, one of the planetesimals in our simulations collides with the Moon rather than with the Earth. For simulations that deliver the late accreted mass to Earth via 1, 2 and 4 planetesimals, the fraction of realizations in which such an event takes place is 9, 15 and 25 percent, respectively. Given the masses that we assume for the planetesimals (0.15-1.2 lunar masses), it is doubtful that the Moon ever experienced such a massive collision. The largest lunar impact for which we have clear and unambiguous evidence is the South Pole Aitken Basin event (a ~$10^{34}$ erg event[27]) that, at the encounter velocities here considered (see Extended

Data Figure 1b), corresponds to a lunar impactor 3-4 orders of magnitude less massive than the planetesimals in our populations. Note that while most basin-forming impacts are thought to have occurred during the late accretion era[17], the effect of these impacts on the lunar inclination was minor[28], (2) Dynamical loss of the Moon. Certain realizations, particularly those that correspond with the strongest tides and largest amount of interacting mass, generate excitations in eccentricity sufficient to destabilize the satellite orbit. Hence, collision with the host planet or (more rarely) liberation of the satellite into heliocentric orbit follows, (3) Too large or too small a mass accreted by the Earth. The total amount of planetesimal mass at the start of simulations was chosen such that, on average, the mass accreted by the Earth would be 0.0075 or 0.015 $M_E$, hereafter denoted the "target mass". Given the stochasticity inherent in this problem, the accreted mass may vary between realizations, resulting in a distribution of outcomes centered on the target mass. To facilitate the expression of the results in terms of late accreted mass, we eliminate from the subsequent analysis those realizations whose accreted mass is greater than or less than the target mass.

Differential momentum transfer is the process underlying this excitation mechanism. For simplicity, we describe this process for the case of a planetesimal colliding with the Earth, but the general case of a collisionless encounter is similar. The orbit of the Earth and Moon around their common center-of-mass is defined by their relative position and velocity. A third body encountering the Earth-Moon binary must have an orbit crossing the system's heliocentric orbit, approaching it with some finite velocity at large separation. Hence, the delivery of mass onto the Earth is accompanied by the delivery of external momentum that – in the impulse approximation – changes the relative velocity but not the relative position of the Earth with respect to the Moon, altering the mutual orbit, a hitherto overlooked effect that can excite the lunar inclination and eccentricity.

We can ask whether the satellite excitation is dominated by the few strongest encounters or the much more numerous distant encounters. Theory suggests that for top-heavy perturber populations, the few strongest perturbations dominate over the more numerous weak ones[29]. We test this for our simulations by generating realizations where the impact parameter is chosen in the interval [0, $R_H$], [0, $R_H/2$] and [0, $R_H/4$], progressively eliminating a large number of distant encounters. The results of the three simulations are statistically indistinguishable (see Extended Data Figure 2a), confirming the theoretical expectation. To facilitate the reproducibility of our results, we plot a measure of the strength of a perturbation versus the impact parameter of the encounters for 2 individual simulations (Extended Data Figure 2b,c).

The simulations are permitted to proceed until the population of Earth-crossing bodies are exhausted either through collision with the Earth or loss via another channel in accordance with an average rate (see above). At the end of the simulations, characteristically lasting ~$10^8$ years, the lunar orbit is typically at ~40 $R_E$. In order to compare the simulation outcomes to the modern system, we propagate the lunar orbit forward to its current separation at 60 $R_E$ and permit the inclination to decay in accordance with the action of planetary tides (Equation M1). The number of simulations was chosen such that the reproducibility of the median inclination values was within several percent.

Recently, it has been suggested that the lunar inclination could have been damped via obliquity tides in the lunar magma ocean (LMO) as the Moon's obliquity increased during its approach to the Cassini state transition between 20-30 $R_E$[19]. The authors of this work put forward one interpretation of the current excited state of the lunar inclination: that the inclination was excited early[12,13] but that the rate of tidal dissipation in the post-giant-impact Earth was sufficiently low as to delay passage of the lunar orbit through the Cassini state transition until after the crystallization of the lunar magma ocean, at which point the effect of obliquity tides on the lunar inclination becomes much less. With the "late" mechanism of lunar orbital excitation here described, we can identify a different solution: that the rate of tidal dissipation on the post-impact Earth is sufficiently rapid in the first tens of millions of years to carry the Moon through the Cassini state transition and to damp any early acquired lunar inclination. Following such a resetting episode, the LMO crystallizes, and inclination excitation due to encounters is subsequently preserved. To explore this solution, we have run a suite of simulations in which the inclination is reset to zero until a certain time, and permitted to accumulate excitation subsequently (see Extended Data Figure 3). Such a transition marks the time of crystallization of the LMO. It can be seen that as long the duration of the lunar magma ocean crystallization is sufficiently short, such a solution is viable and, indeed, necessary in a tidal evolution scenario recently described[30].

**References**
23  Walsh, K. J., Morbidelli, A., Raymond, S. N., O'Brien, D. P. & Mandell, A. M. A low mass for Mars from Jupiter's early gas-driven migration. *Nature* **475**, 206-209, (2011).
24  Wetherill, G. W. Collisions in Asteroid Belt. *Journal of Geophysical Research* **72**, 2429-&, (1967).
25  Yoder, C. F. & Peale, S. J. The Tides of Io. *Icarus* **47**, 1-35, (1981).

26  Kaula, W. M. Tidal Dissipation by Solid Friction and the Resulting Orbital Evolution. *Reviews of Geophysics* **2**, 661-685, (1964).
27  Wieczorek, M. A., Weiss, B. P. & Stewart, S. T. An Impactor Origin for Lunar Magnetic Anomalies. *Science* **335**, 1212-1215, (2012).
28  Canup, R. M. Dynamics of lunar formation. *Annual Review of Astronomy and Astrophysics* **42**, 441-475, (2004).
29  Collins, B. F. & Sari, R. Levy Flights of Binary Orbits Due to Impulsive Encounters. *Astronomical Journal* **136**, 2552-2562, (2008).
30  Zahnle, K. J., Lupu, R., Dobrovolskis, A. & Sleep, N. H. The tethered Moon. *Earth and Planetary Science Letters* **427**, 74-82, (2015).
31  Marchi, S., Bottke, W. F., Kring, D. A. & Morbidelli, A. The onset of the lunar cataclysm as recorded in its ancient crater populations. *Earth and Planetary Science Letters* **325**, 27-38, (2012).

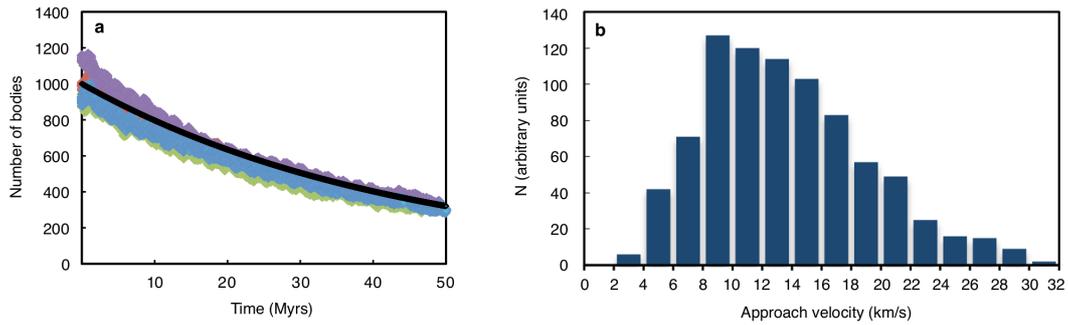

**Extended Data Figure 1 | Properties of planetesimal populations. a,** Decay rates of Earth-crossing planetesimal populations according to N-body simulations of the inner Solar System with a resonant (3:2) Jupiter and Saturn at 5.4 AU and 7.2 AU, respectively. Different colors represent the number of Earth-crossing bodies (hence the evolution not being monotonic) in different simulations, from recent integrations[17]. The black line is the prescribed decay rate used in the three-body simulations ($\tau=44$ Myrs) to match the decay rate in heliocentric simulations. **b,** Histogram of implemented approach velocities (prior to acceleration due to Earth gravity) for late accreting populations in three-body simulations. The population is generated using a Rayleigh distribution of eccentricities and inclinations ($e_R=0.3$, $i_R=e_R/2$) and a semi-major axis range (0.8-1.4 AU) that produces Earth-crossing orbits. These parameters are motivated by simulations of terrestrial planet formation, but the peak of the distribution (9 km/s) corresponds to the typical encounter velocity inferred for lunar basin-forming impactors[31].

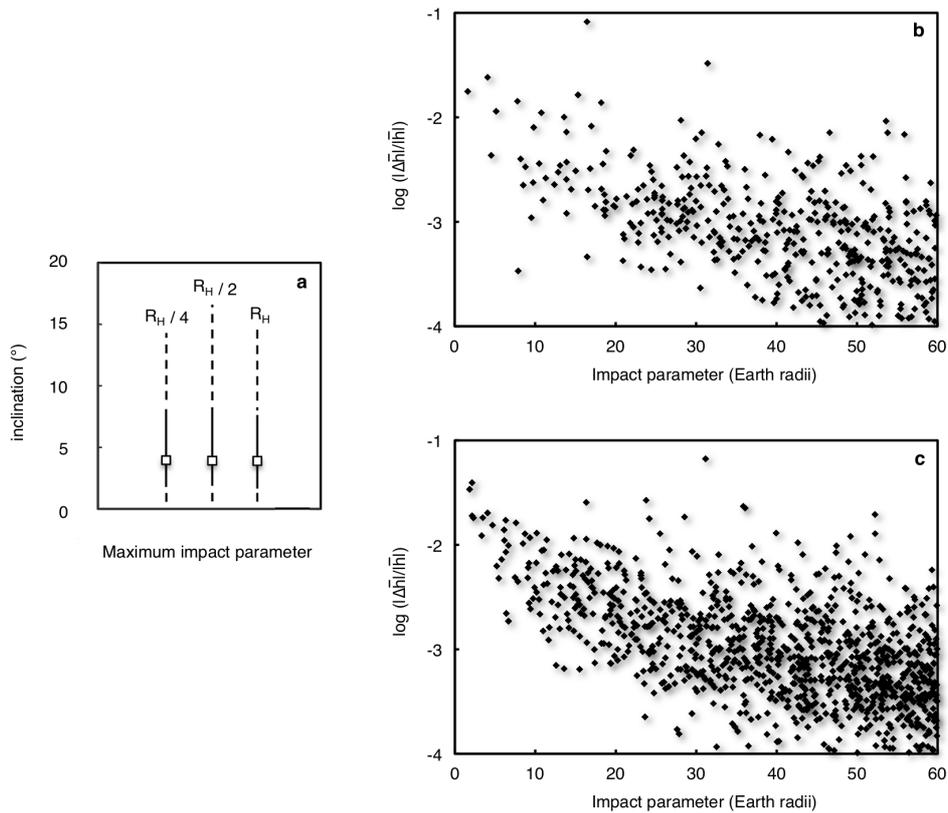

**Extended Data Figure 2 | Tests and outcomes for reproducibility. a,** Test of the cumulative effect of repeated encounters: median values, 1σ (solid), and 2σ (dashed) intervals for three suites of simulations with 0.0075 $M_E$ accreted via a single body onto an Earth with strong dissipation ($k_2/Q$=0.1). Each suite of simulations consists of incoming planetesimals with impact parameters ranging from 0 to $R_H$, $R_H/2$, and $R_H/4$. The statistical identity of the resulting distributions attests to the unimportance of distant encounters as compared to the rare and strong close encounters in generating inclination excitation. **b,c,** Distribution of "kicks" versus impact parameter of encounters from two different realizations. The change in the angular momentum vector of the satellite due to encounter torques is normalized to the magnitude of the *orbital* angular momentum before the encounter. The planetesimals approaching the EM system in this simulation have a mass of 0.0075 $M_E$. Approach velocities are selected from the distribution plotted in Extended Data Figure 1b. The cumulative effects of encounters with b > 60 $R_E$ are negligible and therefore neglected. The top (bottom) panel is from a realization that lasts 26.2 (45.3) Myrs and results in a satellite with a final i=1.9° (8.8°). The strength of tidal dissipation used here ($k_2/Q$=0.1) quickly results in a satellite semi-major axis of 30-40 $R_E$.

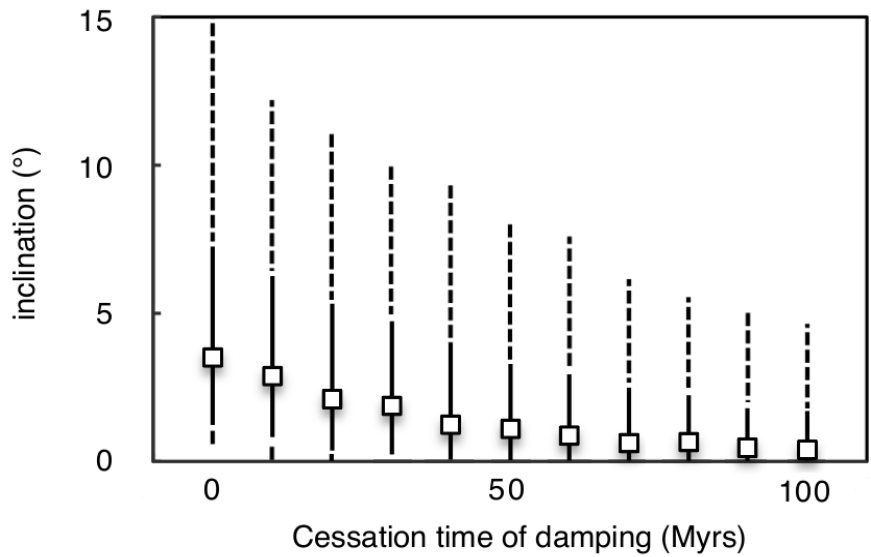

**Extended Data Figure 3 | Effect of partial damping due to LMO obliquity tides.** Median values, 1σ (solid), 2σ (dashed) intervals are shown for several suites of partially damped simulations. These simulations consist of an accreted mass of 0.0075 $M_E$ delivered via a single body onto a strongly dissipative ($k_2/Q=0.1$) Earth, and the orbital excitation continuously reset (e=0, i=0) until a certain time and permitted to accumulate subsequently. Such a transition represents lunar magma ocean crystallization and the cessation of inclination damping via obliquity tides[19]. These simulations are representative of excitation behavior for partially damped cases. It can be seen that, so long as the crystallization of the magma ocean is sufficiently rapid (~$10^7$ years), excitation via planetesimal encounters is largely unaffected.